\documentclass[]{spie}  
\usepackage[]{graphicx}

\title{Future management needs of  a ``software-driven''  \\ science community  }

\author{Kim K. Nilsson\supit{a} and Ole M{\"o}ller-Nilsson\supit{b}
\skiplinehalf
\supit{a} ST-ECF, Karl-Schwarzschild-Stra\ss e 2, 85748, Garching bei M\"unchen, Germany\\
\supit{b} Max-Planck-Institut f{\"u}r Astronomie, K{\"o}nigstuhl 17,
   69117 Heidelberg, Germany
}

\authorinfo{Further author information: \\K.K.N.: E-mail: knilsson@eso.org\\  O.M.-N.: E-mail: moeller@mpia-hd.mpg.de}

\begin{document} 
  \maketitle 

\begin{abstract}
The work of astronomers is getting more complex and advanced as the progress of computer development occurs. With improved computing capabilities and increased data flow, more sophisticated software is required in order to interpret, and fully exploit, astronomic data. However, it is not possible for every astronomer to also be a software specialist. As history has shown, the work of scientists always becomes increasingly specialised, and we here argue in favour of another, at least partial, split between ``programmers'' and ``interpreters''. In this presentation we outline our vision for a new approach and symbiosis between software specialists and scientists, and present its advantages along with a simple test case.
\end{abstract}



\section{INTRODUCTION}
\label{sec:intro}  
The work of a present day scientist is vastly different from that of the early scientists such as Leonardo da Vinci, Galileo Galilei and Tycho Brahe. These renaissance scientists were polymaths, studying vastly different subjects such as astronomy, mathematics, music, botany and anatomy, to mention a few. What makes present day scientists different? For one, they have access to extremely more detailed instruments and observations. They also have access to several hundreds of years of knowledge accrued, and in several fields of study the largest questions have already been answered and these answers need only be refined. But one ultimate difference between present day scientists, compared to any scientist working merely 50 years ago is the use of computers to facilitate data analyses. It is on the basis of this evolution in the scientific process that scientists become more and more specialised on a sub-topic within a larger field of study, and why scientists now have to be just as proficient in the methods of programming as they are in coming up with theories, devising new experiments and returning new results. 

Astronomers do extremely diverse work. Not only do they spend time analysing and interpreting results, the main scientific process furthering our knowledge, but they also write proposals and scientific publications, prepare observing runs and give presentations. But most of all, they work on producing the scientific results by means of computers and programming; both on the basic level of data reduction and pipeline writing, as well as on higher level data exploration such as photometric redshift codes, spectral energy distribution (SED) fitting codes, morphological modeling codes etc. And this includes only observational astronomers. Astronomers working on producing theoretical models obviously spend a much larger part of their day simply in writing software. 

But astronomers are not particularly trained to write good code. Many have no formal training in programming, and few stay up to date with what new languages and techniques are developed over the years. Computer scientists on the other hand have no formal understanding of the scientific process, or how to conduct an astronomical experiment. That is why we think the time is ripe for a synergetic link to be formed between the two professions. Time has shown that astronomers get more and more specialised within their fields. But why then cling to the idea that researchers have to write all their software themselves? In this publication, we suggest a split between ``programmers'' and ``interpreters''. With this split, where major programming tasks are completed by the software specialist, the researchers save a large amount of time on devising new tests and analysing their results, i.e. to ``interpret'' the data. 

Our approach is to first investigate if such a new role of astronomical software specialist is motivated from the perspective of the software community. To this end we devised a questionnaire that we sent out to several European astronomy institutes.
We present the questionnaire and the results in sec.\ref{sec:data}. We then discuss in more detail under what conditions the hiring of a such a software specialist increases the productivity more than hiring a science focused astronomer would. We present some simple arguments for why this would make sense even for relatively small institutes in sec.~\ref{sec:testcase}. We also propose and describe a pilot test we envision that could be used to confirm the positive effects of this collaborative relationship. Section \ref{sec:newcareer} discusses in more detail the educational aspects and likely impacts on the work and social patterns within the astronomical community. We also attempt to predict possible stumbling blocks in introducing this "new type of astronomer" and outline what we believe can be learned from existing attempts to introduce software support into astronomical and related research. Finally we conclude in sec.~\ref{sec:conclusion}.

\section{QUESTIONNAIRE: QUESTIONS AND  RESULTS}
\label{sec:data}
In order to verify the interest and need for special software specialists at astronomy institute, an online questionnaire was distributed to several (mostly European) institutes. The analysis here is based on 142 individual answers from astronomers at varying stages of their career. Roughly a third each of the group were students, post-docs/fellows or staff. The majority were male ($70$\%) and their ages were distributed in 39\% younger than 30, 36\% between 31 and 40, 18\% between 41 and 50 and 6\% above 50 years old. The fact that the age groups exhibit a pyramidal distribution, whereas the percentage of respondents are equal in the three career steps indicate that there is not a clear correlation between age and career step. We also conclude that we are biased towards the answers of younger astronomers age-wise, who are presumably more likely to answer an online questionnaire. This bias is not entirely negative though, since the young astronomers will stay longer at the astronomy institutes and thus represent the future needs of the community.

\begin{figure}
   \begin{center}
   \begin{tabular}{c}
   \includegraphics[height=5cm]{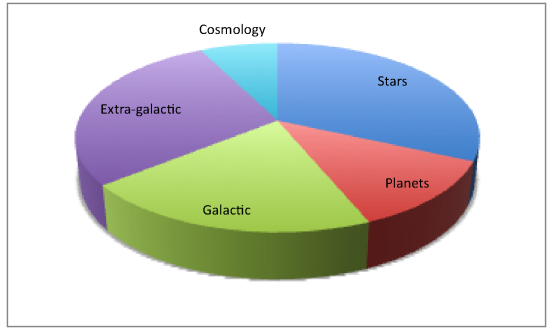}
   \end{tabular}
   \end{center}
   \caption[example] 
   { \label{fig:topics} 
Distribution of topics among the astronomers who answered the questionnaire.}
   \end{figure}

\begin{figure}
   \begin{center}
   \begin{tabular}{c}
   \includegraphics[height=5cm]{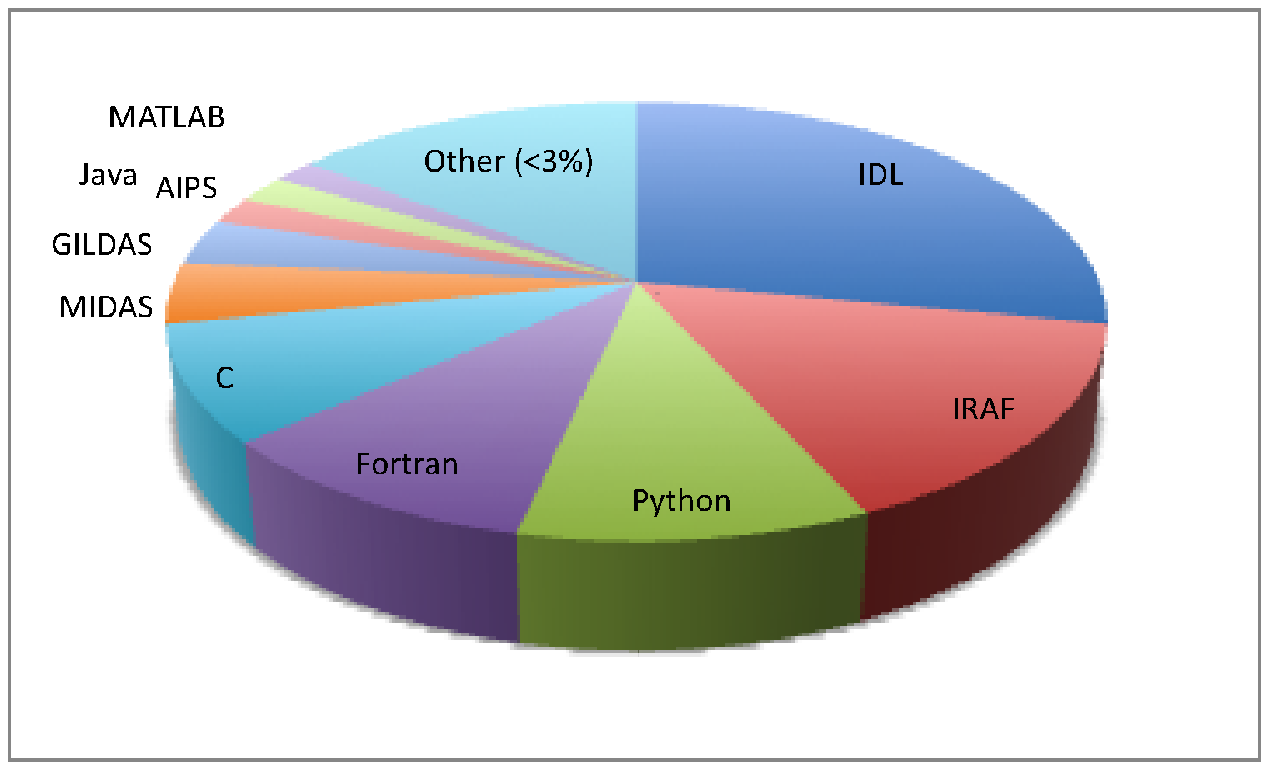} \includegraphics[height=5cm]{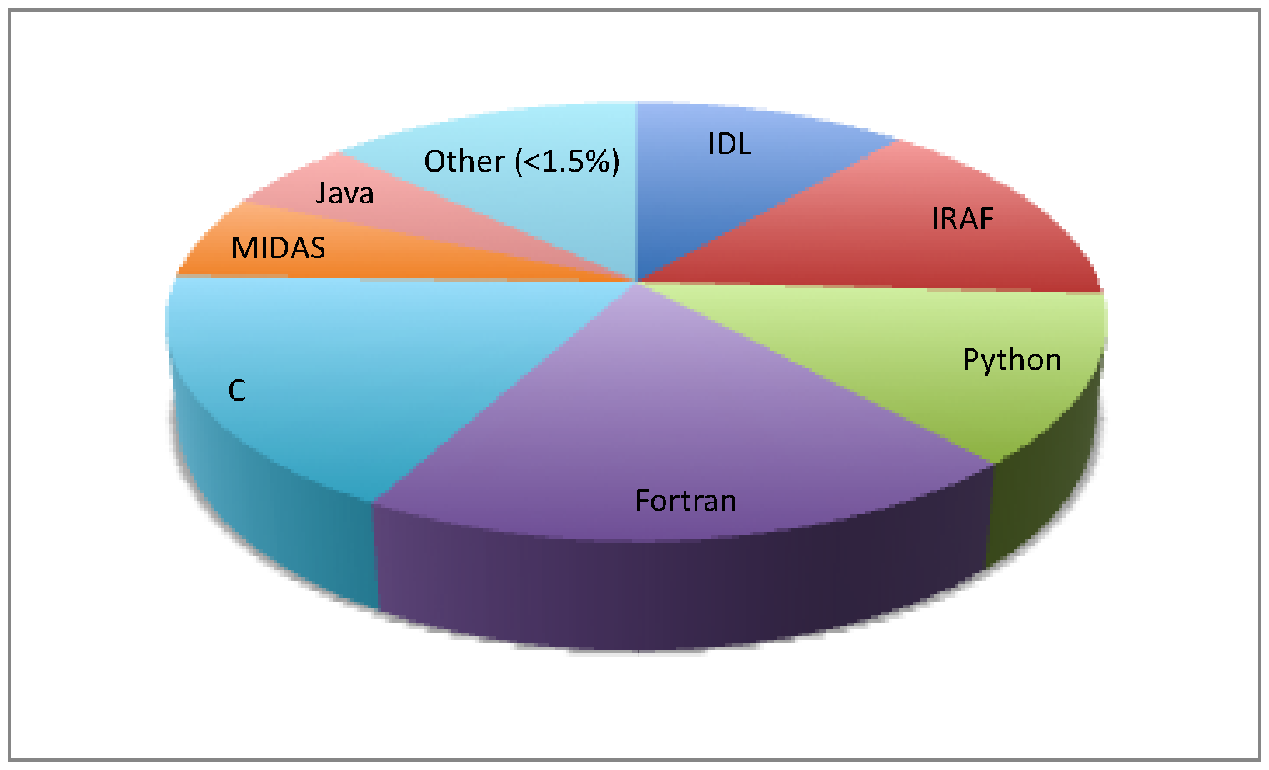}
   \end{tabular}
   \end{center}
   \caption[example] 
   { \label{fig:langs} 
Distribution of main programming languages (\emph{left}) and occasional programming languages (\emph{right}) used by the astronomers. The numbers from this plot can be found in Table~\ref{tab:langs}.}
   \end{figure}

Among the basic information asked for was also the topic of research. Initially, $28$\% replied they were working with theoretical models, $80$\% were observers and $22$\% worked with instrumentation. The topics of research were divided into the groups of \emph{i)} stars and star formation, \emph{ii)} planets and planet formation, \emph{iii)} galactic astronomy, \emph{iv)} extra-galactic astronomy and \emph{v)} cosmology. The resulting answers can be found in Fig.~\ref{fig:topics}.
Further, the respondents were asked to mark which programming languages they used regularly (i.e. several times weekly) and occasionally (a few times a month). The result can be seen in Fig.~\ref{fig:langs} and in Table~\ref{tab:langs}.
\begin{table}[h]
\caption{Main and occasionally used programming languages by astronomers. The ``Other'' category includes all languages with less than $3$\% of users in the main category and less than $1.5$\% in the occasional category (including awk, bash, CASA, cuda, Excel, FTOOLS, GDL, GIPSY, gnuplot, HEASOFT, HIPE, hyperz, KARMA, Labview, mathematica, MIR, MIRIAD, Paw, PHP, PDL, Perl, PhoS-T, R, SAS, shell, SQL, Starlink, STILTS,  supermongo, tcl, tk, Yorick, wip).  } 
\label{tab:langs}
\begin{center}       
\begin{tabular}{|l|l|l|l|} 
\hline
\hline
\rule[-1ex]{0pt}{3.5ex}   & Main & Occasional & Total  \\
\rule[-1ex]{0pt}{3.5ex}   & \% & \% & \% \\
\hline
\rule[-1ex]{0pt}{3.5ex}  IDL & 57.7 & 17.6 & 75.3  \\
\rule[-1ex]{0pt}{3.5ex}  IRAF &  31.0 & 23.2 & 54.2 \\
\rule[-1ex]{0pt}{3.5ex} Fortran & 20.4 & 31.0 & 51.4 \\
\rule[-1ex]{0pt}{3.5ex}  C/C++ & 17.6 & 27.5 & 45.1 \\
\rule[-1ex]{0pt}{3.5ex}  Python & 21.1 & 19.7 & 40.8 \\
\rule[-1ex]{0pt}{3.5ex}  MIDAS & 9.2 & 9.9 & 19.1 \\
\rule[-1ex]{0pt}{3.5ex}  Java & 4.2 & 8.5 & 12.7 \\
\rule[-1ex]{0pt}{3.5ex}  GILDAS & 7.0 & $<1.5$ & $<8.5$ \\
\rule[-1ex]{0pt}{3.5ex}  AIPS & 4.2 & $<1.5$ & $<5.7$  \\
\rule[-1ex]{0pt}{3.5ex}  Matlab & 3.5 & $<1.5$ & $<5.0$ \\
\rule[-1ex]{0pt}{3.5ex}  Other & 29.6 & 20.4 & 50.0 \\
\hline 
\end{tabular}
\end{center}
\end{table} 
The main languages used by astronomers are IDL, IRAF and Fortran which are all used by more than $50$\% of all respondents either on a weekly or monthly base. Closely following comes Python, C/C++ and MIDAS. The perhaps greatest surprise is the multitude of programming languages used regularly by few users. Almost $50$\% of all respondents use one of 33 languages regularly, but these languages only have a few percent of users each. 

Another question was posed regarding ``high end data products'' such as fully reduced images or spectra, source catalogues, redshift catalogues, simulations etc. Of the astronomers who answered the questionnaire, $12$\% had no use of these data products in their science. Another $3$\% answered that they did not trust high end data products and never used them. Of the remaining group, there were varying levels of trust in public data varying from those who test all products before using them ($17$\%) to those that trust some and test some, depending on source, ($63$\%) and those that always trust the data products in full ($5$\%). The answers indicate what seems to be a skepticism among astronomers to trust data prepared by other groups, with $80$\% of all astronomers testing some or all of the high end data products they use.

The main question of the questionnaire was that of whether the respondent would make use of the service of a software specialist. The options for an answer where \emph{i)} no, never, \emph{ii)} yes, for learning about existing software, \emph{iii)} yes, for minor help with programming issues and bugs and \emph{iv)} yes, for help with writing major software projects. Several of the positive answers could be marked. Of the respondents, the vast majority ($91$\%) marked that they would appreciate the help of a software specialist. Of these, 57\% would want help with existing software, 50\% would like help with minor programming issues and 56\% would like to work in parallel with a software specialist on major projects. 60\% said that the quality of their work would improve with the help of a software specialist and of the 48\% who answered the question of how much time they would save per week for scientific interpretation, the average time saved was 5.6 hours per week. This is a fraction of 14\% of the total working week (assuming a 40 hour working week), a number we will call the $X_s$ for fraction saved. If we, for a conservative estimate, assume that all those in favour of working with a software specialist that did not specify a time saved would save no time per week, the average time spent on programming becomes 2.7 hours per person and week. This gives a conservative fraction of $X_s = 6.7$\%. It is also worth noting that, as expected, the percentage of respondents who said they had no use for a software specialist was significantly larger among the theoreticians ($15$\%) than among observers ($6$\%).

\section{BENEFITS OF A SPECIALIST}
\label{sec:testcase}

Based on the results of our questionnaire, we see that there is a great need for interactions between astronomers and software specialists. The number of hours potentially saved for scientific interpretation indicates that most astronomy institutes could be made much more efficient if help was provided with small and large programming projects. However, since the number of hours saved was a pure estimate by the respondents, it is clear that this needs to be tested. It is also clear that there would be a need for a test phase, in order to determine what projects such a software specialist could help with, what other tasks this person could be given (e.g. educational) and the actual efficiency achieved. We thus provide a suggestion for a pilot programme in this section.

\subsection{Specialist tasks}
Assume an institute will hire a software specialist to support the scientific projects, what questions need to be answered? What should the background of such a specialist be? What would his/her tasks be? What programming languages does this expert need to be familiar with? How much actual time will be saved for scientific interpretation by the astronomers helped?

To start with the first question, the seemingly ideal background of an astronomy software specialist would be a degree in astronomy. It is crucial that the specialist understands the special challenges in astronomical research, and the tools to solve them. Some further training in programming and computer science would of course be desirable. We discuss this subject further in Sec.~\ref{sec:newcareer}. The languages that the specialist needs to be familiar with are dictated by the results of the questionnaire, as shown in Fig.~\ref{fig:langs} and in Table~\ref{tab:langs}. The diversity in languages used by astronomers make it difficult to master all possibilities, but the specialist should be familiar with the main languages IDL, IRAF, Fortran, C and Python. Ideally, the specialist should also have deeper knowledge in at least two of these. As for the tasks, we divided the potential tasks into three groups in the questionnaire. The most passive task would be that of keeping an inventory on what astronomical software is publicly available, and advice on the use of these programs. However, note that this specialist is not meant to be a general IT helpdesk or technical support. Secondly, the specialist could offer minor assistance with coding, such as bug searching, code optimisation techniques etc. In these tasks we also potentially foresee some form of teaching capability regarding programming and languages, individually or in groups. Finally, each specialist could take on larger projects, in close collaboration with the scientists, writing specially designed code to solve a particular question that the scientists want answered. An example of such a project would be that of writing an SED fitting code in the most efficient and accurate way, based on the scientists particular data-set. Obviously the exact tasks would need to be pre-defined, also partly based on the exact background and preferences of the specialist. 

\subsection{Efficiency considerations}
One might ask the question; ``At what point does it increase the scientific output more for an institute to hire a software specialist, rather than one more astronomer''? The answer is simply that as long as the time saved for scientific interpretation is larger than zero, and as long as the software specialist can be kept busy full-time, then hiring a software specialist to relieve the scientists of some programming burden will \underline{always} increase the scientific efficiency of the institute more than hiring another astronomer. The simple reason is because as long as the software specialist is kept busy, even assuming that the programming efficiency of the specialist is equal to that of the astronomer which is a conservative estimate, this person will deliver the equivalent of a $100$\% work week of pure scientific productivity compared to an astronomer who is not $100$\% efficient. Basically the total number of hours spent on scientific output before the new hiring is
\begin{equation}\label{eq1}
N \times ( 1 - X_s ) \times h
\end{equation}
where $N$ is the number of astronomers at the institute, $X_s$ is the fraction of time spent on tasks that could be done better by a software specialist and $h$ is the number of working hours per week. If another astronomer is hired, the total number of hours worked will be
\begin{equation}\label{eq2}
( N + 1 ) \times ( 1 - X_s ) \times h
\end{equation}
However, if a software specialist is hired, taking over the fraction of programming time needed until the working hours are full, and assuming the same programming efficiency of astronomers and specialists, the total scientific output will be
\begin{equation}\label{eq3}
( N - \frac{1}{X_s} ) \times ( 1 - X_s ) \times h + \frac{1}{X_s} \times h
\end{equation} 
We see that if Eq.~\ref{eq3} is larger than Eq.~\ref{eq2}, then it makes more sense to hire a software specialist rather than an astronomer. When these equations are worked out, it is clear that this condition is met when $X_s > 0$, i.e. as long as the working hours saved is larger than zero. This is of course only true as long as there are not too many specialists compared to astronomers. With some common sense reasoning, this is true when
\begin{equation}
N - N_s \times \frac{1}{X_s} > \frac{1}{X_s}
\end{equation}
where $N_s$ is now the number of software specialists already hired at the institute. This can be simplified to
\begin{equation}
N \times X_s - N_s > 1
\end{equation}
or, inserting the number from the questionnaire results,
\begin{equation}
N \times [ 0.067 - 0.14 ] - N_s > 1
\end{equation}
In conclusion, one software specialist on $7-15$ astronomers should maximise the efficiency of an astronomy institute.

Several assumptions went in to the calculation above. First, the number of hours saved by each astronomer each week if helped by a software specialist is uncertain and could be wrongly estimated in both positive and negative direction. Second, it seems reasonable to assume that a software specialist can complete the same programming task in less time than the astronomer, although we above assume a similar efficiency for a conservative estimate. On the other hand, time will also be spent on the interaction between the specialist and the astronomer, time which is difficult to estimate. Thus the need for a pilot programme.

\subsection{Pilot programme}
A pilot project could go on for a year, and should be extensively evaluated during, and after the end of, the project. The time saved (or the increased efficiency of the institute) needs to be confirmed, and the best way would be for the test subjects (both scientists and the specialist) to keep a brief diary on how much time was spent on programming issues, and how much time was saved. It is also important for both the astronomers and the specialist to maintain a list of positive and negative feed-back regarding the interactions between the groups, and how this could be improved. After a year, the pilot programme can be evaluated and a recommendation made regarding future hiring of software specialists at the institute. Note that the software specialist has to be in all aspects regarded as an integral part of the scientific work carried out at the institute. He/she must not be seen as a "second grade" scientist, and should be encouraged to take active part in the scientific life of the institute. It may also be desirable to encourage co-authorship of papers that present research carried out using software developed/selected by the software specialist.

\section{DISCUSSION}
\label{sec:newcareer}
\begin{figure}
   \begin{center}
   \begin{tabular}{c}
   \includegraphics[height=5cm]{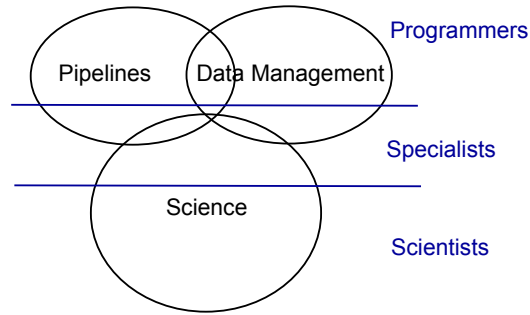}
   \end{tabular}
   \end{center}
   \caption[example] 
   { \label{fig:work} 
Areas of software expertise involved in scientific research within observational astronomy. Both data management and data processing (pipelines) are needed to ``feed'' the scientific work. The scientific analysis that derives results to scientific questions overlaps partly with the other areas, and it is in this area that a significant amount of software is written by astronomers. A software specialist would take over mainly those tasks that fall in these overlap areas.}
   \end{figure}

Programming in (observational) astronomical research is roughly divided into three groups, illustrated in Fig.~\ref{fig:work}.
The three areas are data management (e.g.~handling of large databases), data reduction (or ``pipelines'') and scientific analysis and deduction. Currently, the most common situation in astronomy is that the first two areas are covered largely by programmers (without a science background) and the last is covered by a scientist who develops software only for his/her own needs. The tasks of the software specialist that we suggest here would be located somewhere in the middle of Fig.~\ref{fig:work}, providing an interface between the purely computational programming tasks of writing pipelines and managing large databases, and the scientific programming. We want to explicitly acknowledge here that there are several research institutes where the employment of software developers, similar to what we propose here, is a fact. However, we point out that this is in most cases not scientifically motivated, but motivated purely by instrument control, data management or data reduction. Software engineers are currently hired as part of instrument development teams, for the sole purpose of making large amount of data available to the scientific community, either in a raw or partly reduced format, or creating tools that reduce astronomical data (data reduction pipelines, the two top areas in Fig.~\ref{fig:work}). Most of this work can in principle be carried out by software engineers without any astronomical background. Software engineers that are employed to be directly involved in the scientific research carried out at an institute hardly exist.

It is here instructive to also look at other disciplines for comparison. In particle physics especially is there a vast amount of software development connected with experiments and theoretical research. The situation there is driven by the need for complex data management of a vast amount of data. This has lead to a rather extensive but distinct software development group that develop and support a large collection of software tools that allow data analysis. In essence this is not different to the data management and analysis provided as part of large public astronomical surveys, such as SDSS. This type of software is of course very important, but it is again different from the scientifically motivated software development and support we envision here. This is again illustrated by the fact that the software developers for these tools are in general not physicists.

A possible problem for the introduction of software specialists is the lack of adequate candidates. The currently existing positions for software engineers at astronomical institutes can rarely be filled with people that have both research experience in astronomy and experience in software engineering on a professional level. We believe, however, that this is related to the fact that there is currently no "software engineering culture" in astronomy. Software development is regarded as far less important than paper publication. Astronomers that have a natural desire and skill in software development are often pushed out into industry. They see no clear career path within astronomy where they can both be actively involved in research and yet spend a significant amount of time on developing good software, and furthering their knowledge of software engineering. Given how strongly the personal publication rate affects the chances of a permanent position today, it is all but unavoidable that software development can never be more than a pure means for personal paper production. Writing high quality software that can be shared freely requires a significant time and effort that reduces publication rate dramatically. 

In the long term the kind of scheme we propose here is thus only feasible if there is a change in the hiring policies in the astronomical community. Since a software specialist will have a much lower publication rate (which may include no first author papers) the software specialist cannot be hired in direct competition with other research astronomers. Programming must become a clear, distinct new career path for astronomy PhD students. This needs to be advertised, and options must exist already at the PhD stage to include significant educational modules in programming and software engineering aspects. Positions for software specialists would then be filled by applicants with a PhD in astronomy and strong skills in, and enthusiasm for, software engineering.

\section{CONCLUSIONS}
\label{sec:conclusion}

Based on personal experiences, we set out to present an idea for increasing the scientific efficiency of astronomy institutes by hiring dedicated software specialists to help the astronomers with the more technical aspects of their work. These specialists would represent a third class of hires at an institute, next to scientists and programmers, interfacing between pure programming projects and science projects. In order to gauge the interest in the community for such a service, a questionnaire was launched and answered by $>140$ active astronomers. The response was overwhelming. $90$\% of all astronomers would appreciate help with programming tasks, saving somewhere between $2.5 - 5.5$ hours a week for scientific interpretations. If this is confirmed, an institute could in principle increase its scientific efficiency by up to $14$\% by hiring software specialists. In Sec.~\ref{sec:testcase} we outlined a pilot programme lasting one year that can in principle be adopted at any astronomy institute and that will result in a more accurate estimate of the efficiency increase of the institute. We have also noted the importance of a clear career path for students who are interested in astronomy, but whose strengths lie in programming, rather than science. We believe this is the future for astronomy. As computers become more advanced, so do the programs. Theoretical models become more detailed and observing instruments more complex requiring more complicated data reduction pipelines. But astronomers also get more and more specialised in their sub-field of astronomy and may then loose overview of the tools and techniques of other sub-disciplines. This is not negative trend, but rather a natural evolution of science, and the scientific process. The only question for each institute to answer is whether it is fighting against this evolution, or leading it.

\acknowledgments     
 
The authors are grateful for the $\sim150$ answers received by scientists on the questionnaire regarding their software habits and requirements.


\bibliography{report}   
\bibliographystyle{spiebib}   

\end{document}